\documentclass{aa}
\usepackage[varg]{txfonts}

\usepackage{graphicx}
\usepackage{subfigure}
\usepackage{amssymb}
\usepackage[english]{babel}
\usepackage[varg]{txfonts}
\usepackage{chngpage}
\usepackage{natbib}
\usepackage{threeparttable}
\usepackage{pgffor}
\usepackage{tikz}
\usepackage{lipsum}
\usepackage{rotating}
\usepackage{geometry}
\usepackage{lscape}
\usepackage{longtable}
\usepackage{appendix}
\usepackage{comment}

\def\vycma{VY~CMa\xspace}
\def\cit6{CIT~6\xspace}

\def\nmlcyg{NML~Cyg\xspace}

\def\dfk{DFK~52\xspace}

\def\kms{km\,s$^{-1}$\xspace}
\def\msunyr{$M_{\odot}$\,yr$^{-1}$\xspace}
\def\msun{$M_{\odot}$\xspace}
\def\lsun{$L_{\odot}$\xspace}

\def\aap{A\&A}

\def\apjs{ApJS}
\def\apjl{ApJL}

\begin{document}

\title{Stephenson~2~\dfk: Discovery of an exotic red supergiant in the massive stellar cluster RSGC2}

\author{M. A. Siebert\inst{1}
  \and E. De Beck\inst{1}
  \and G. Quintana-Lacaci\inst{2}
  \and W. Vlemmings\inst{1}
  }

\institute{Department of Space, Earth and Environment, Chalmers University of Technology, Gothenburg,  Sweden 
 		 \and Department of Molecular Astrophysics, Instituto de F\'isica Fundamental, Madrid, Spain
         }

\date{Accepted July 6, 2025}

\abstract 
{Atacama Large Millimeter/submillimeter Array (ALMA) observations at 1.3mm have recently revealed surprising complexity in the circumstellar environment of \dfk, a red supergiant (RSG) located in the Stephenson 2 massive open cluster. We provide an initial  characterisation of the star's mass-loss properties by studying its circumstellar emission in continuum, $^{12}$CO, $^{13}$CO, and SiO rotational lines. We find that \dfk is surrounded by an extremely large outflow (up to 50,000\,au in radius) that shows complex morphologies in both its molecular and dust emission. The size of the circumstellar medium is unprecedented, even when compared with other known extreme RSGs, and its  lower luminosity indicates that its mass ejection mechanism may be unique among this population. The molecular emission can be partially reproduced by a two-component model consisting of a fast (27\,\kms) detached equatorial component with $M{\sim}0.05$\,\msun and a slow (10\,\kms) spherical envelope with $\dot{M}\sim3\times10^{-6}$\,\msunyr. This suggests that \dfk underwent a dramatic mass-loss event $\sim$4000 years ago, but has since transitioned into having a slower  more symmetric mass loss. We conservatively estimate a total mass of $0.1-1$\,\msun in the complex extended regions of the outflow. The uncertain nature of the dramatic mass loss warrants extensive follow-up of this likely supernova progenitor. }

\keywords{ Stars: supergiants, Stars: mass-loss, Stars: circumstellar matter,  Stars: evolution}
\maketitle 

\nolinenumbers

\section{Introduction}\label{sect:intro}
The mass loss of red supergiants (RSGs) is critical to their evolution and to the observed characteristics of Type II-p supernovae (SNe). The efficiency of these winds for the duration of the RSG phase can determine the amount of hydrogen stripped from the star, the density of interacting dust at the time of explosion, and the structure of SN remnants. Among well-studied RSGs, a morphological distinction has emerged between those exhibiting slower, traditional, approximately spherical mass loss (e.g. $\alpha$ Ori, $\alpha$ Sco; $10^{-7}-10^{-6}$\msunyr) and more luminous extreme RSGs, which show asymmetric outflows and mass-loss rates of order $10^{-4}$\msunyr (e.g.\ VY CMa, NML Cyg, VX Sgr). Neither the underlying physical mechanisms that produce these exotic objects nor their ubiquity among the red supergiant population is understood, despite their major relevance to evolutionary models of massive stars and core-collapse SNe.

Stephenson 2 (RSGC2) is a massive cluster of at least 26 RSGs located at the base of the Scutum-Crux spiral arm at a distance of $5.8^{+1.91}_{-0.76}$\,kpc \citep{davies2007_rsgc2}. It has been inferred that this was a site of recent starburst activity in the region where the arm intersects the galactic bulge. \citet{humphreys2020} investigated the mass-loss properties of cluster members from dust emission in their mid-infrared (IR) spectral energy distributions (SEDs), and noted its wide spread in luminosities, initial masses, and mass-loss rates when compared to other RSG clusters. 

In these previous studies, the source \dfk appears as a rather ordinary RSG in the cluster. Its radial velocity and plane-of-sky position place it in the centre of RSGC2, the strength of its CO absorption band in the near-IR implies an M0-type supergiant \citep{davies2007_rsgc2}, and its estimated luminosity ($2\times10^4$\,\lsun) is typical, though on the low end for RSGs. However,  \citet{humphreys2020}  noted that its $J$-band flux is too low to be reproduced with a stellar model and the average reddening factor for the cluster. This could either arise from variable line-of-sight extinction from the intervening interstellar medium (ISM), or from circumstellar dust.

In this letter we report the discovery of a surprising circumstellar environment toward \dfk based on recent observations with ALMA. Through its molecular line and continuum emission at millimetre wavelengths, we present the kinematics and spatial distribution of gas and dust surrounding this star, and compare them with other known extreme RSGs. With these constraints, we aim to provide starting insights into the evolutionary status and the mass-loss history of \dfk that will serve as a basis for future more detailed studies of this unique object.

\section{Observations}\label{sect:obs}

The millimetre-wavelength observations of DFK~52 were conducted as part of two ALMA programmes. The first (2023.1.01519.S) covered a sample of 13 RSGs in RSGC2. These observations were performed with two configurations of the main 12m array using the Band 6 receiver \citep{Ediss2004_Band6}, achieving a spatial resolution of 0.35\arcsec\/ at 230\,GHz and a maximum recoverable scale (MRS) of 12.5\arcsec. Four spectral windows were positioned in the $214-233$\,GHz range targeting $^{12}$CO $J=2-1$ at 230.538\,GHz and SiO $J=5-4$ at 215.596\,GHz. The spectral resolution is 1.27\,\kms, and the total bandwidth is 1.8\,GHz. The follow-up project (2024.A.00018.S) targeted only \dfk with the Atacama Compact Array (ACA) to recover all of the extended emission. The spectral set-up for the ACA matches the resolution and line coverage of the 12m data, with the addition of the $^{13}$CO $J=2-1$ transition at 220.399\,GHz.

Observations for both ALMA projects were reduced with the standard ALMA calibration pipeline \citep{Hunter2023_ALMApipeline}, and visibilities from different configurations were combined in the \textit{uv}-plane with antenna-specific weighting. Continuum subtraction was performed in the visibility plane, and imaging was done with the CASA task \texttt{tclean} using Briggs weighting and a robust parameter of 0.5. The final brightness sensitivities are 1\,mJy/beam for spectral line data and 20\,$\mu$Jy/beam for the continuum. A common feature across all ALMA observations of RSGC2 was a significant amount of contamination from the interstellar medium (ISM) in CO. To reduce its impact on our results, we identified the channels where ISM emission or absorption spatially overlaps with circumstellar CO, and did not consider them in our analysis. For continuum maps, line-free channels were used in the imaging process, and the $^{12}$CO spectral window was omitted entirely to ensure the ISM was excluded. Comparing images created with and without the supplementary ACA data, the total flux recovery is ${\sim}100$\% for spectral lines and 80\% for the continuum in the main-array dataset.

\section{Analysis and results}\label{sect:analysis_results}
\subsection{Continuum}\label{sect:cont}

\begin{figure}[t!]
    \centering
    \includegraphics[width=.9\linewidth]{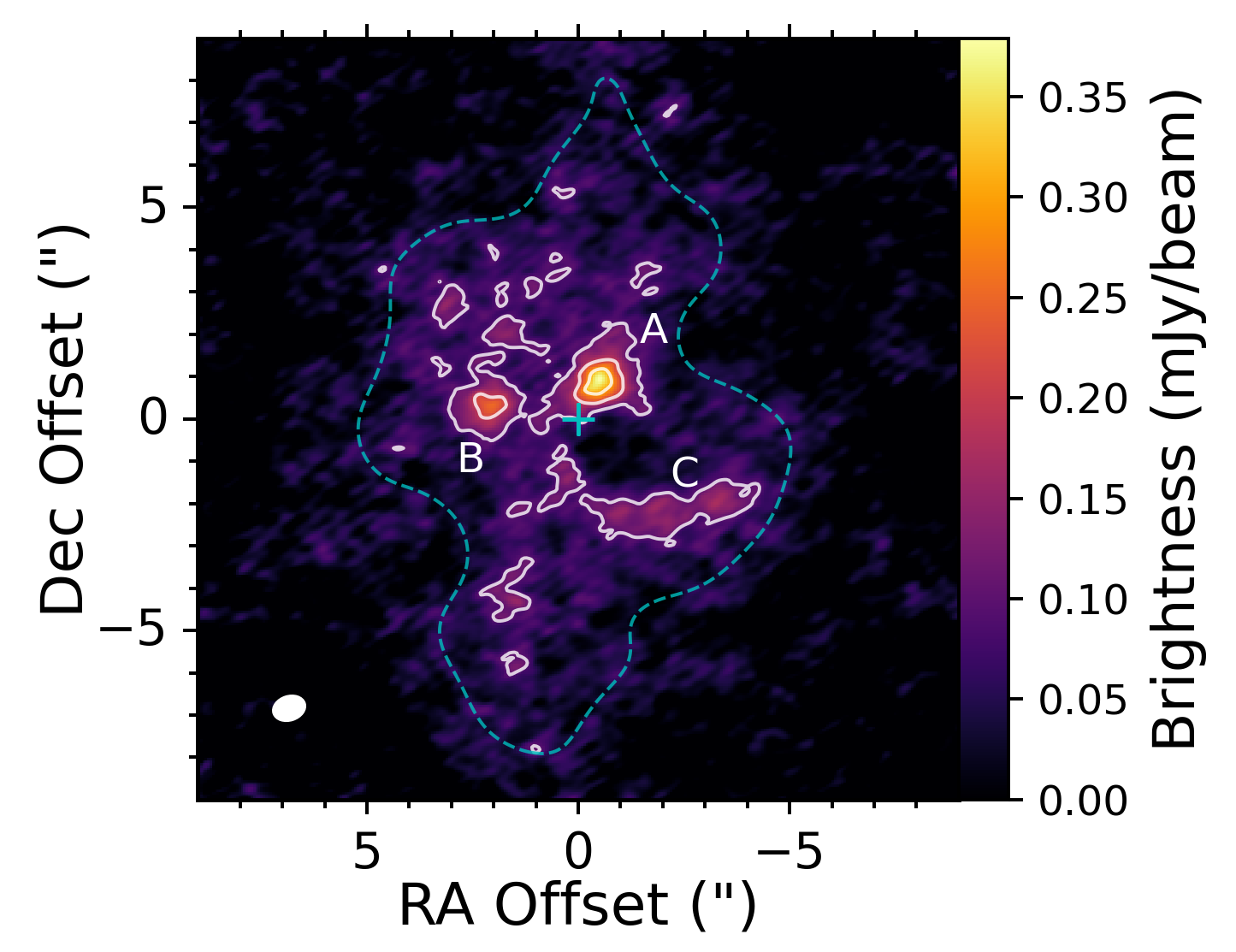} 
    \caption{220\,GHz continuum toward \dfk. The stellar position (see Section \ref{sect:co}) is marked with a blue cross, and the A, B, and C components are labelled. The white contours are shown at $[5,10,15]\sigma$, and the dashed blue contour represents $3\sigma$ for the image residuals after smoothing to a 2\arcsec\/ resolution, tracing the low-brightness extended component.}
    \label{fig:continuum}
\end{figure}

The 1.3\,mm continuum toward DFK~52 is shown in Fig.~\ref{fig:continuum}. We observed emission in a variety of substructures, including a very low-brightness (${\sim}1$-$3\sigma$ unsmoothed) extended component that is elongated in the N--S direction and reaches a maximum radius of 8\arcsec. At shorter radii, more compact components are present, including two bright clumps in the NW and E directions (labelled A and B), and an arc structure to the SW (labelled C). 
Together, these structures account for ${\sim}$15\% of the total flux. Notably, no centrally peaked continuum source is found at the position of the star (measured by unresolved SiO emission; Section \ref{sect:co}). The total continuum flux is $17.2\pm1.7$\,mJy. 

The continuum map appears similarly complex in shape to the extreme RSG \vycma, where the millimetre emission is dominated by thermal radiation from asymmetrically ejected dust clumps \citep{kaminski_vycma_2019}. However, the surrounding \vycma are observed at projected distances of $250-800$\,au \citep[see clumps `A--D' in ][]{Humphreys2024_VYCMa_clumps}, whereas the A and B clumps around \dfk are found at $5,800$ and $12,000$\,au, and the extended emission reaches a distance of $\sim50,000$\,au (assuming the cluster distance of 5.8\,kpc; see Appendix \ref{app:membership}). Following the methods of \citet{ogorman2015} and \citet{debeck2025_nmlcyg} for \vycma and \nmlcyg and using a dust grain emissivity of $\beta=0.9$, we derive dust temperatures $T_{\mathrm{A}}\approx102\pm9$\,K, $T_{\mathrm{B}}\approx76\pm7$\,K, and $T_{\mathrm{C}}\approx63\pm6$\,K, and dust mass estimates  $M_{\mathrm{A}}\approx(1.7\pm0.36)\times10^{-4}\,M_{\odot}$,  $M_{\mathrm{B}}\approx(1.1\pm0.24)\times10^{-4}\,M_{\odot}$, and $M_{\mathrm{C}}\approx(4.4\pm0.95)\times10^{-4}\,M_{\odot}$. This assumes that the A, B, and C continuum components are situated at offsets of
$\Delta_{\mathrm{A}}\approx1.0^{\prime\prime}$, 
$\Delta_{\mathrm{B}}\approx2.1^{\prime\prime}$, and 
$\Delta_{\mathrm{C}}\approx3.2^{\prime\prime}$
in the plane of the sky and that their emission is optically thin. These masses are comparable to the estimates for continuum dust clumps around VY CMa, which have $M\sim10^{-4}$\,\msun \citep{Humphreys_2022}.

Adding to the peculiarity of the continuum map, the spectral energy distribution (SED) shows a clear double-peaked shape (Appendix \ref{fig:sed}). This is different from \vycma and \nmlcyg, whose SEDs exhibit much higher fluxes in the mid-IR contributed by significantly warmer dust found close to the star \citep{kaminski_vycma_2019,Justtanont1996_NMLCyg}. Considering the large physical scales in the ALMA image, the secondary peak in the SED, and the lack of mid-IR excess or millimetre emission at the stellar position, we suggest that the dust emission is dominated by a cold detached component. Using the same method as before, but for the total ALMA flux, and adopting an average dust temperature of 50\,K (inferred from the far-IR SED peak). This gives $M_{\mathrm{d}}\approx(6.6\pm0.7)\times10^{-3}\,M_{\odot}$, or a total envelope mass of $1.3$\,\msun if a gas-to-dust ratio of 200 is assumed \citep{humphreys2020}.

\begin{figure*}[t!]
    \centering
    \subfigure[CO $J=2-1$ and SiO $J=5-4$ (inset) moment-zero maps.]{\includegraphics[trim={-1cm 0cm -1cm 0cm},clip,height=0.42\linewidth]{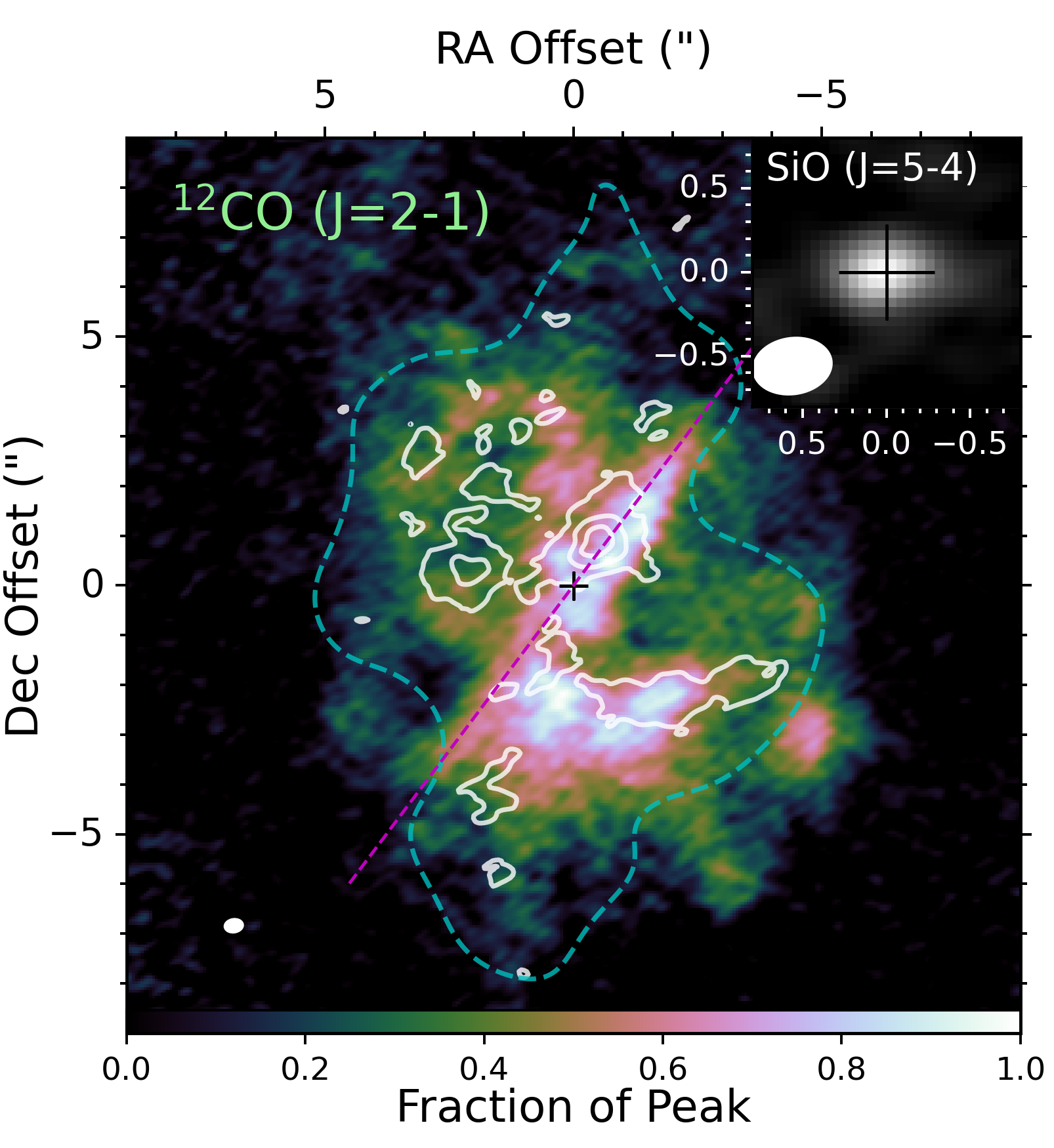} \label{fig:co_mom0}}
    \hspace{3em}
    \subfigure[CO isotopologue (top) and SiO (bottom) spectra. ]{\includegraphics[trim={-1cm 0cm -1cm 0cm},clip,height=0.42\linewidth]{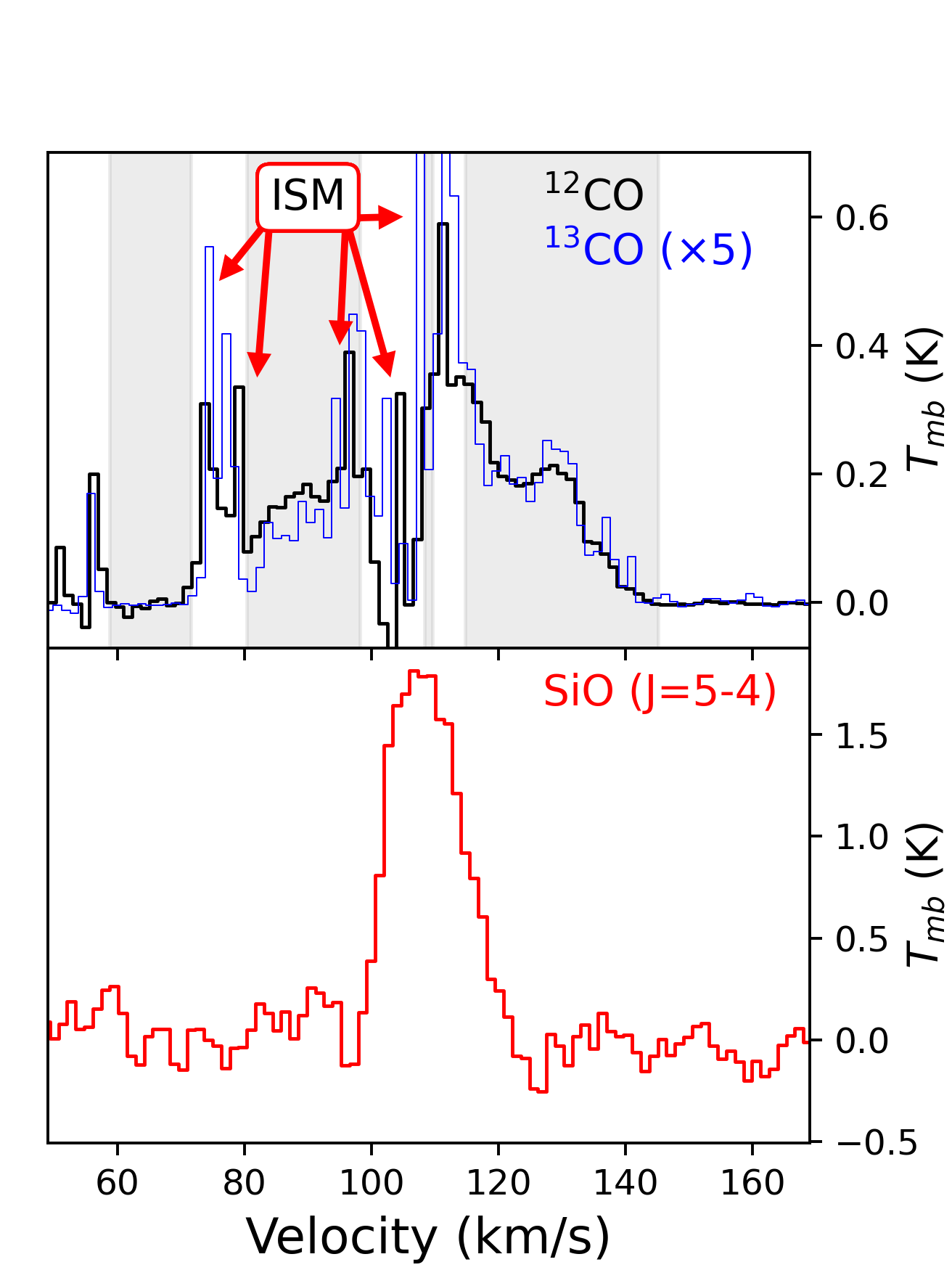} \label{fig:co_sio_spectra}}
    \caption{Summary of molecular line emission observed toward DFK52. In the integrated CO map, the white contours denote the 220\,GHz continuum emission at $[5,10,15]\sigma$, the blue dashed contours represent $3\sigma$ for the smoothed extended continuum component, the black cross marks the stellar position (as measured by the SiO emission peak), and the ALMA beam is displayed as a white filled ellipse in the lower right corner. The dashed magenta line in the integrated intensity map represents a line cut with $PA=143^{\circ}$. Spectra were extracted taking the central pixel at the stellar position, with CO isotopologue maps first smoothed to a 15\arcsec\/ beam and SiO maps at the native resolution. The grey shaded regions in the CO spectrum indicate the ISM-free channels that were used to produce the moment-zero $^{12}$CO image.}
    \label{fig:morph_summary}
\end{figure*}

\subsection{Molecular emission}\label{sect:co}
The spatial and kinematic morphology of rotational line emission seen toward \dfk is summarised in Fig.\ \ref{fig:morph_summary}. We briefly present a number of important features.

\begin{itemize}
    \item SiO $J=5-4$ is spatially unresolved, and parabolic in line shape. This is consistent with optically thick thermal emission at the stellar position, as is measured toward many AGB and RSG stars. From this, we measure the systemic velocity $v_{sys}=109\pm1.2$\,\kms and stellar position $\alpha_{\mathrm{J2000}}=18\mathrm{h}39\mathrm{m}23.406\mathrm{s}$, $\delta_{\mathrm{J2000}}=-06^{\circ}02'16.12''$.
    \item With a maximum radius of $\sim$7.5\arcsec, the integrated $^{12}$CO emission is significantly more extended than SiO. Moreover, the CO line width of $\Delta v_{10\%}\approx54$\,\kms traces gas that moves $\sim$2.5 times faster than SiO ($\Delta v_{10\%}\approx20$\,\kms).
    \item The CO morphology is highly asymmetric, containing numerous arcs, loops, and roughly linear components. We identify a bar-like structure extending along position angle $\mathrm{PA}=143^{\circ}$ and across the stellar position. No local maximum is measured at the stellar position.
     
    \item The size and orientation of the integrated CO emission are in quite good agreement with the extended continuum. The C component is traced well by $^{12}$CO channel maps in the velocity range $130-138$\,\kms, while the southward continuum extension is seen in the range $92.5-98$\,\kms (Appendix \ref{app:chmaps}).
    
    \item Despite these correlations, the A and B dust clumps have no counterparts in CO. This may result from the  missing  channels contaminated by the ISM. We also find that some CO flux extends beyond the scales of the diffuse dust emission.

    \item The physical radial extent of $^{12}$CO $J=2-1$ structures is 45,000\,au. This is roughly four times larger than the farthest components observed in this line for VY CMa \citep["NE Arc";][]{singh2023} and NML Cyg \citep{debeck2025_nmlcyg}.
\end{itemize}

\begin{figure}[t!]
    \centering
    \includegraphics[width=.9\linewidth]{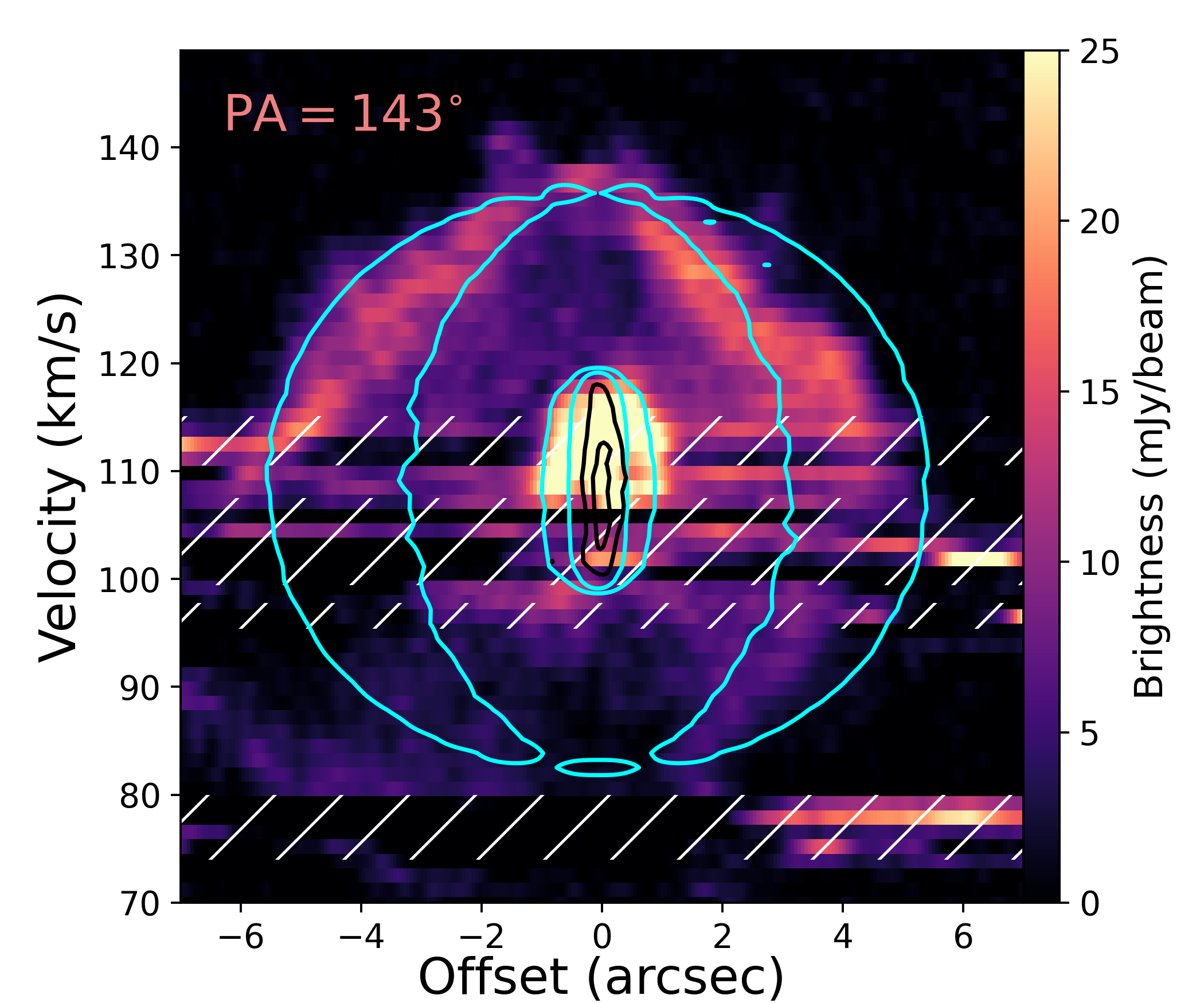}
    \caption{PV diagram of $^{12}$CO ($J=2-1$) emission for a cut at $PA=143^{\circ}$ with a 1\arcsec\/ width. The black contours show SiO ($J=5-4$) at $3\sigma$ and $10\sigma$. The blue contours are simulated $^{12}$CO ($J=2-1$) emission from our two-component RT model, shown at 5 and 15 mJy/beam. The hatched regions denote velocity ranges heavily affected by ISM contamination.}
    \label{fig:model_summary}
\end{figure}

Though no simple geometric prescriptions can explain the overall structure, the bar-like component in the integrated CO map (Fig.\ \ref{fig:co_mom0}) shows a significant amount of coherence across velocity space. Figure \ref{fig:model_summary} shows the position-velocity (PV) diagram for $^{12}$CO taken along the bar, where a shell shape is clearly seen extending to $v_{\mathrm{sys}}\pm27$\,\kms. This pattern suggests that the bar represents a detached disk-like structure viewed edge-on. The PV diagram also shows a bright smaller component with a total velocity width of $\sim$20\,\kms, consistent with the SiO $J=5-4$ emission. This inner structure is also seen in a perpendicular cut, suggesting that it has more 3D extent than the bar.

To estimate their physical properties, we  reproduced the two components outlined above using the radiative transfer (RT) software LIne Modeling Engine (LIME 1.9.5) \citep{Brinch2010_LIME}. The model parametrises the two structures as a slow spherically symmetric present-day mass loss, and a faster detached equatorial density enhancement (EDE) viewed edge-on (Appendix \ref{app:lime}). A model with an equatorial mass of 0.05\,\msun centred at $r=4$\arcsec\, (4000\,yr age) and a central component with $\dot{M}=3\times10^{-6}$\,\msunyr is successful at reproducing the average surface brightness and radial distribution of emission both in the compact and extended regions along the position angle of the bar structure (Fig.\ \ref{fig:model_summary}). Furthermore, the derived mass-loss rate for the slow component is in agreement with the SED modelling of warm dust performed by \citet[][where a gas-to-dust ratio of 200 was adopted]{humphreys2020}. However, some shortcomings of the model are clear as well. First, the blueshifted side of the torus emission is largely missing from the observational data. No combination of optical depth and excitation conditions could reproduce this, so we conclude that there is an azimuthal density asymmetry in this structure. Furthermore, there is a substantial amount of CO emission between the two components that is not captured by the model.

Because the LIME model was constructed to fit only a small fraction of the observed outflow at a specific position angle, it reproduces 15\% of the total line flux. Although the extended emission outside the position angle of the bar is extremely complex, we note that its velocity distribution is still quite similar to the EDE component (Appendix \ref{app:lime}). If we assume the excitation conditions of CO are common between the many extended density structures, then the model can be scaled to provide a very rough estimate of the cumulative mass in the detached high-velocity components surrounding \dfk. Doing so offers a total mass of ${\sim}$0.35\,\msun, which is within a factor of four agreement with the estimate obtained from the continuum flux in Section \ref{sect:cont}.

\section{Discussion}\label{sect:discussion}
\dfk is a puzzling object for several reasons. The estimated physical extent of its molecular envelope is a factor of $3-4$ larger than the outflows of \vycma and \nmlcyg \citep{singh2023,debeck2025_nmlcyg}, and many of the irregular geometries have no clear analogues in any observed evolved star. At the same time, the luminosity is a factor of 10 lower than the known extreme RSGs \citep{humphreys2020}, and its present-day mass loss is estimated to be two orders of magnitude smaller. We find that both the molecular line and continuum emission require a significant mass of cold material at large radii ($>10^{17}$\,cm) in the form of complex detached structures. 

Given the high stellar density in the RSGC2 cluster and the strong ISM contamination in these observations, it is possible that the large-scale emission represents interactions between a stellar outflow and a quiescent local density structure in the intra-cluster medium. The maximum physical extent of the envelope is comparable to the smallest observed galactic clumps \citep[0.1\,pc;][]{Uruquhart2021_ISM_ATLASGAL}, and the smooth underlying continuum emission is more similar in size and shape to molecular cloud substructures than circumstellar envelopes \citep{Tokuda2019_LMC_filaments}. However, comparison of the $^{13}$CO and $^{12}$CO emission across the high signal-to-noise ratio channels free of ISM contamination leads to $^{12}$CO/$^{13}$CO$\approx5$ (Fig.\ \ref{fig:co_sio_spectra}). This low ratio supports a stellar origin of the presented molecular emission as it requires stellar processing through the H-burning via the CNO cycle and dredge-up of the material to the surface \citep{Boothroyd1993_HBB}. We also note that none of the other sources in the RSGC2 sample show morphology anywhere similar to this, supporting the idea that the morphological complexity is connected to the star rather than to a complex intra-cluster component.

To explain the light curves of Type II SNe, it has been suggested that RSGs explode into a very dense immediate circumstellar medium (CSM), built up by a superwind ($>10^{-3}$\,\msunyr) in the $\sim$100 years prior to core collapse \citep{Moriya2017_RSGsuperwinds,davies2022}. The CO emission around \dfk indicates that a period of dramatic mass loss with $\dot{M}>10^{-4}$\,\msunyr occurred $\sim4000$ years ago. If we assume constant radial expansion for the disk component in the RT model, this portion of the CSM would have had a peak gas density of $n\sim10^{10}$cm$^{-3}$, or $\rho\sim10^{-14}$g\,cm$^{-3}$ when it was at a distance of $10^{14}$\,cm from \dfk, which is comparable to the pre-SN CSM density required by \citet{Yaron2017_SN2013fs}, for example. However, our analysis also suggests that this period was immediately followed by an ongoing slower mass loss with $\dot{M}\sim3\times10^{-6}$\,\msunyr, and that the central gas density is lower at present. Therefore, if \dfk did experience a superwind, it ended without producing a SN.

Alternatively, an energetic interaction in a binary or multiple system could have contributed to the ejection of the extended material. Companion interactions and stellar mergers are known to produce bipolar, multipolar, equatorial, and complex morphologies \citep[e.g.][]{Lagadec2011_pAGB_mIRimaging,olofsson2019,Kaminski2018_rnovae_submm}, and common envelope evolution (CEE) is now considered to be widely responsible for rapid mass-loss events observed in AGB and post-AGB sources \citep{Khouri2021_CEEs}. Furthermore, CEE in RSG systems has been invoked as an important channel to the progenitors of ultra-stripped SNe, as well as the compact binaries ($P<1$\,day) responsible for gravitational wave events \citep{Wei2024_RSG_CEEmodels,Klencki2021_RSG_BH_CE}. CE ejections are known to produce expanding equatorial rings similar to the partial one seen in \dfk \citep{Gomez2018_I15103}, but given the lack of bipolar symmetry and the absence of any high-velocity jet components ($>30$\,\kms), this assignment cannot be made with the current observations. We still consider a multiple-star hypothesis to be strong here, as \dfk's comparably low luminosity would have required an additional source of energy to drive the massive detached components (Appendix \ref{app:momentum}).

\section{Conclusion}\label{sect:conclusion}
We have presented resolved emission from the circumstellar environment of the red supergiant \dfk, revealing a dusty envelope spanning unprecedented physical scales. Both the continuum and CO imply the presence of a massive ($0.1\sim1$\,\msun) asymmetric, equatorially enhanced, detached wind expanding at $v\approx27$\,\kms. Additionally, compact emission near the stellar position implies that the mass-loss rate and expansion velocity have dropped significantly since the ejection of this component. With the available data, a full classification of \dfk and its history is not yet feasible; however, we consider a short-lived superwind period or an energetic companion interaction to be strong possibilities that explain the circumstellar structure. The exotic nature of \dfk makes it a useful candidate to study major mass-loss events that can occur during the RSG phase, so we emphasise the need for future multi-wavelength studies of this object.

\begin{acknowledgements}
We thank the reviewer for their helpful comments on this letter. This paper makes use of the following ALMA data: ADS/JAO.ALMA\#2023.1.01519, ADS/JAO.ALMA\#2024.A.00018.S. ALMA is a partnership of ESO (representing its member states), NSF (USA) and NINS (Japan), together with NRC (Canada), NSTC and ASIAA (Taiwan), and KASI (Republic of Korea), in cooperation with the Republic of Chile. The Joint ALMA Observatory is operated by ESO, AUI/NRAO and NAOJ. We gratefully acknowledge the use of Director's Discretionary Time at ALMA under project ADS/JAO.ALMA\#2024.A.00018.S. G.Q. would like to thank funding support from Spanish Ministerio de Ciencia, Innovación, y Universidades through grant PID2023-147545NB-I00.
\end{acknowledgements}

\bibliographystyle{aa} 
\bibliography{ref.bib}

\begin{appendix}
    \section{Spectral energy distribution}
    \label{app:sed}
\begin{figure}[h]
        \centering
        \includegraphics[width=\linewidth]{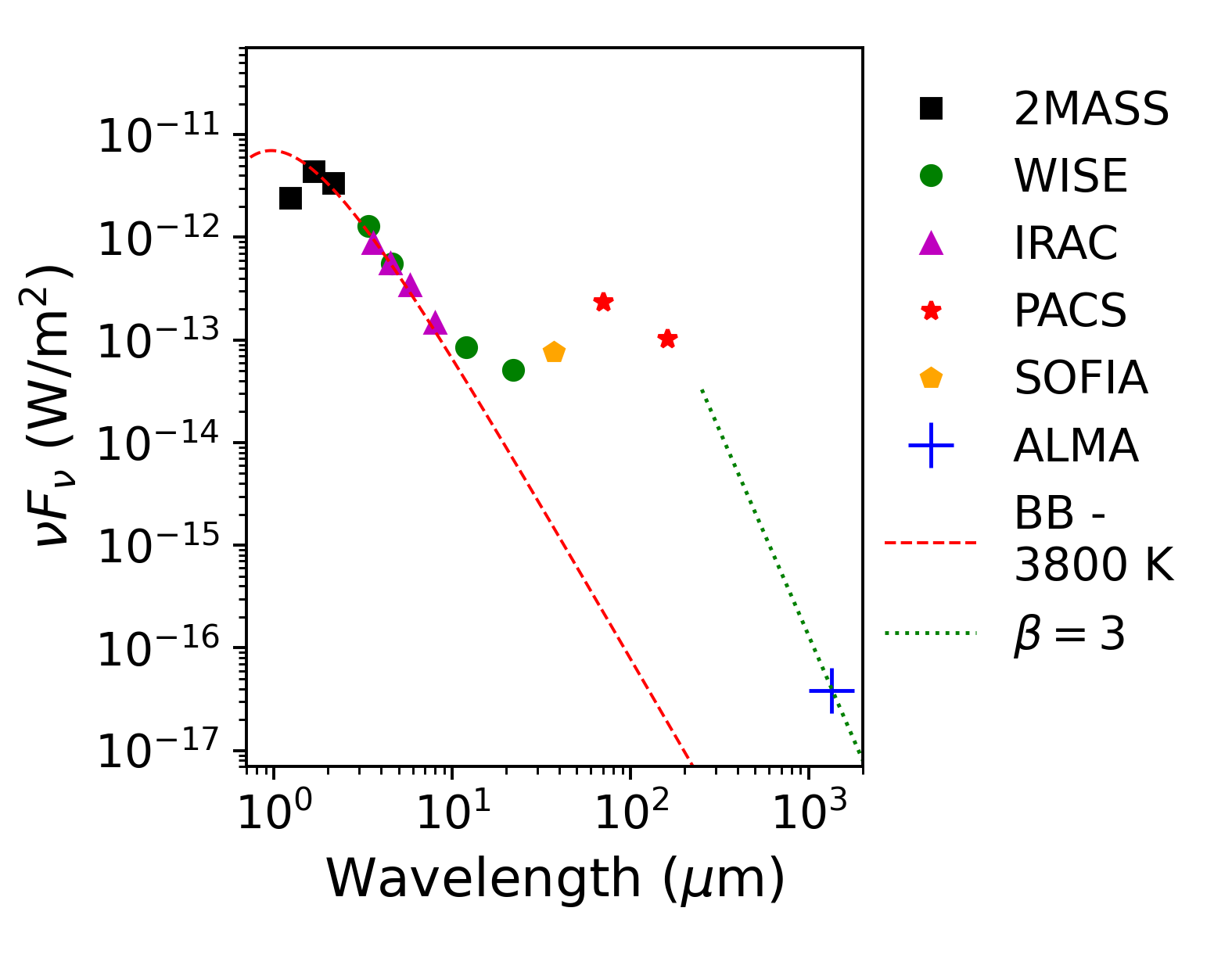}
        \caption{SED for \dfk, including total ALMA flux measured in this work. Archival photometric measurements were corrected for foreground interstellar extinction by the same amounts as \citet{humphreys2020}. The lines are shown denoting the stellar blackbody (dashed red) as well as a spectral index of 3 extrapolated from the ALMA point (dotted green).}
        \label{fig:sed}
    \end{figure}

    We added the new ALMA 1.3\,mm continuum flux and archival measurements from the Herschel/PACS point source catalogue \citep{Marton2024_PACS_PSC} to the photometry compiled by \citet{humphreys2020} to construct the SED for \dfk shown in (Fig.~\ref{fig:sed}). This reveals a double-peaked shape, with a cold dust component maximum at $\sim70\,\mu$m, yielding a temperature of $\sim50$\,K. Although the spectral indices of the individual continuum components are unavailable with the current ALMA observations, a linear interpolation from the total ALMA flux to the PACS 120\,$\mu$m yields an estimate of $\alpha\approx3$, implying optically thin dust grains \citep{ogorman2015} and supporting this assumption made in our dust estimates in Section \ref{sect:cont}.

    \section{Details of the 3D radiative transfer model}
    \label{app:lime}
    In the two-component LIME model, the central component is treated as a spherically symmetric expanding wind with constant mass-loss rate. The extended component, similar to the set-up of \citet{Sahai2022_VHya}, is modelled as an equatorial density enhancement (EDE) described as a 2D Gaussian in (spherical) radius and the $z$-coordinate:
    \begin{equation}
        n(r,z) \propto \exp\left(\frac{-\left(r-r_{\mathrm{EDE}}\right)^2}{2\sigma_{\mathrm{EDE}}^2}\right)\times\exp\left(\frac{-z^2}{2h^2}\right).
    \end{equation}
Here $r_{\mathrm{EDE}}$ is the radius where the density peaks, $\sigma_{\mathrm{EDE}}$ is the standard deviation of the enhancement, and the vertical scale height $h$ is determined by the cylindrical radius coordinate $R$ and an opening angle, $\theta_{\mathrm{EDE}}$:
    \begin{equation}
        h(R) = R\tan\left(\theta_{\mathrm{EDE}}/2\right).
    \end{equation}
    The temperature profile for both structures is a power law in radius with log-slope $\alpha=1$, based on the stellar temperature and radius. A minimum kinetic temperature, $T_{\mathrm{min}}=15$\,K, is also needed because this power law reaches unphysically low values at large radii where the torus is found.
    
    Due to the very different radial density dependences of the two components, populations were simulated separately using $\log r$ grid-sampling for the central component, and uniform sampling for the equatorial component. The excitation is collisionally dominated, and the observations indicate that the components do not overlap spatially at any velocity, so we do not expect the central and equatorial components to interfere with each other in optical depth or rotational populations. The parameters of the final RT model are included in Table \ref{tab:lime_params}. 

\begin{figure}[h!]
    \centering
    \includegraphics[width=\linewidth]{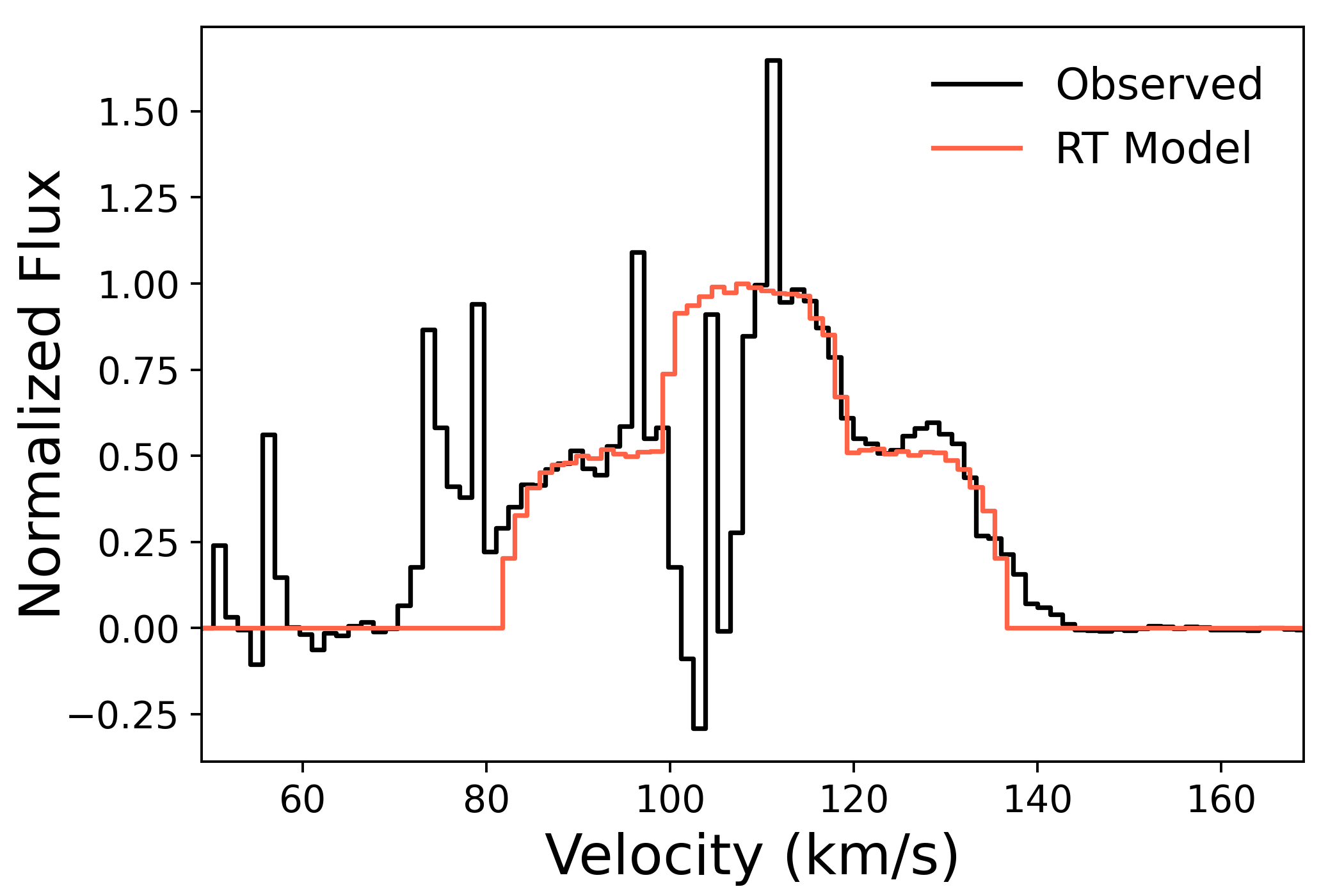}
    \caption{Observed $^{12}$CO $J=2-1$ line profile (black) and output of the two-component model (red). We note that normalised lines are used here to compare the shape of the emission, and that the RT model only accounts for 15\% of the total observed flux.}
     \label{fig:lime_speccomp}
    \end{figure}

    A comparison between the observed and modelled normalised line profile shapes is shown in Fig. \ref{fig:lime_speccomp}. We find that the fast EDE component and slow central component scale quite well to the total emission. While this suggests that material outside the equatorial regions may be comparable in its velocity distribution and optical depth properties, a full 3D model of the extended material is not yet possible due to the very high degree of geometrical complexity.

    \begin{table}[h!]
    \centering
    \begin{tabular}{l c}
    \hline\hline                        
        Parameter & Value \\    
        \hline
        $^{12}$CO abundance w.r.t.\ H$_2$ & $2\times10^{-4}$ \\ 
        Stellar temperature, $T_\star$ & 3800\,K $^a$ \\
        Stellar luminosity, $L_\star$ & $2.4\times10^{4}$\,\lsun $^a$ \\
        Temperature, $T_{\mathrm{kin}}$ & $T_\star\left(\frac{r}{R_\star}\right)^{-\alpha}$ \\
        \hline 
        \textbf{Central component} & \\
        Mass-loss rate, $\dot{M}$ & $3\times10^{-6}$\,\msunyr \\
        Expansion velocity, $v_{\mathrm{sph}}$ & $10$\,\kms \\
        \hline
        \textbf{Equatorial component} & \\
        Total mass, $M_{\mathrm{EDE}}$ & 0.05\,\msun \\
        Expansion velocity, $v_{\mathrm{EDE}}$ & 27\,\kms\\
        Center radius, $r_{\mathrm{EDE}}$ & 23000\,au \\
        Radial width, $\sigma_{\mathrm{EDE}}$ & 6000\,au \\
        Opening angle, $\theta_{\mathrm{EDE}}$ & 10$^{\circ}$ \\
        \hline\\
    \end{tabular}
    \caption{Physical parameters of the two-component RT model. $^a$Adapted from \citep{humphreys2020}.}
    \label{tab:lime_params}
    \end{table}
   
    \section{Equatorial momentum analysis}
    \label{app:momentum}
    
    The linear momentum estimated for the equatorial outflow, based on the derived parameters (see Table~\ref{tab:lime_params}), is approximately \(2.7 \times 10^{38}\,\mathrm{g\,cm\,s^{-1}}\). To evaluate whether radiation pressure could account for this value, we computed the momentum it can deliver over the acceleration timescale using the formulation by \citet{val2001}, \(P = F \cdot P_{\mathrm{rad}} = F \cdot L \cdot t_{\mathrm{acc}} / c\), where the efficiency factor \( F \) describes the ability of dust to absorb and re-emit photons multiple times, enhancing momentum transfer. Assuming the current luminosity of \(2 \times 10^4\,L_\odot\) and an acceleration time of 1000 years---which corresponds to the duration of the ejection of the EDE---the resulting radiative momentum is \(P_{\mathrm{rad}} \approx 8.05 \times 10^{37}\,\mathrm{g\,cm\,s^{-1}}\) and the implied efficiency factor is \(F \approx 3.35\).

    In oxygen-rich massive post-AGB stars, the efficiency of momentum transfer from radiation pressure to the circumstellar gas is expected to be lower than in carbon-rich environments. This is primarily due to the reduced formation of carbonaceous dust grains, which are highly efficient at absorbing stellar radiation. Instead, oxygen-rich stars predominantly form silicate-based dust, which has lower opacity in the infrared, and thus couples less effectively with the radiation field. In environments with lower dust-to-gas ratios or less efficient dust species, this process is less effective and thus leads to reduced values of \( F \). Additionally, in equatorial outflows, the geometry further limits the fraction of stellar radiation that can be redirected into the plane of the flow, so the \( F \) calculated above for \dfk must be considered a lower limit. As a rough estimate, for a torus height of 4000\,au ($\approx 10^{\circ}$) and a radius of 23000\,au, the EDE could intercept only about 9\% of the total stellar radiation.

    \citet{val2001} scaled the value of the effective opacity with the mass loss; their value of \( F \) is expected for mass losses of $\sim 10^{-4}$\msunyr. The average mass loss we obtained for the EDE is $\sim 5 \times 10^{-5}$\msunyr (=0.05\msun/1000\,yr). Thus, if we expect only 9\% of the total of the photons to reach the region of the EDE and a $\tau_\mathrm{eff}\,\sim 5$ times lower, we should reach (even assuming the same dust characteristics)  a value of \emph{F} of $9-17 \times 10^{-2}$, significantly smaller than the minimum efficiency derived. 
    This supports the conclusion that an alternate ejection mechanism must be invoked to explain the formation of the EDE observed. If the equatorial component was indeed accelerated by radiation pressure, the stellar luminosity would have had to be 1--2 orders of magnitude higher than it is estimated now. Such a dramatic drop in luminosity over a timescale of a few $10^3$\,yr  is not consistent with any predictions of single-star RSG evolutionary models \citep[e.g.][]{Zapartas2025_Mdot_RSGmodels}; it is, however, consistent with the timescales and stellar parameter changes modelled for some merger products \citep[e.g.][]{schneider2020_merger}.

    \section{Cluster membership}
    \label{app:membership}
    The interpretation of these data is reliant on the assumption that \dfk is a cluster member, and hence at the cluster distance of $5.8^{+1.91}_{-0.76}$\,kpc \citep{davies2007_rsgc2}. If \dfk were in the background, its luminosity could be more in line with other extreme RSGs, but the size of the gas and dust emitting regions would cover even larger spatial scales than currently derived, further complicating the question of how such an envelope can be formed. On the other hand, if \dfk were in the foreground, the physical scales of emission would decrease; however, its luminosity would be even lower than currently quoted, challenging the star's classification as an RSG. In this case, the star would have a much later spectral type, due to the empirical relationship between $T_{\mathrm{eff}}$ and CO band head strength for non-supergiants \citep{GonzalezFernandez2012_RSGC3,davies2007_rsgc2}, and because the corrected SED would be substantially more reddened \citep{negueruela2012}. A very evolved foreground AGB star could be considered, but even at a distance of 1.5\,kpc (where the envelope size would be comparable to VY CMa), the luminosity would then be small even for AGBs ($<2000$\,\lsun). The systemic velocity ($109$\,km/s; measured with SiO) also lies comfortably within the cluster velocity range ($100-120$\,km/s), and would be quite peculiar for a $d<2$\,kpc star, as the galactic rotation curve in this sight line predicts $v_{\mathrm{lsr}}<30$\kms in that case. For these reasons, we argue for the current assignment that DFK~52 is indeed a cluster member, but a more accurate distance measurement is needed to confirm this.

    \section{$^{12}$CO channel maps}
    \label{app:chmaps}
    
    \begin{figure*}[t!]
    \centering
    \includegraphics[width=.90\linewidth]{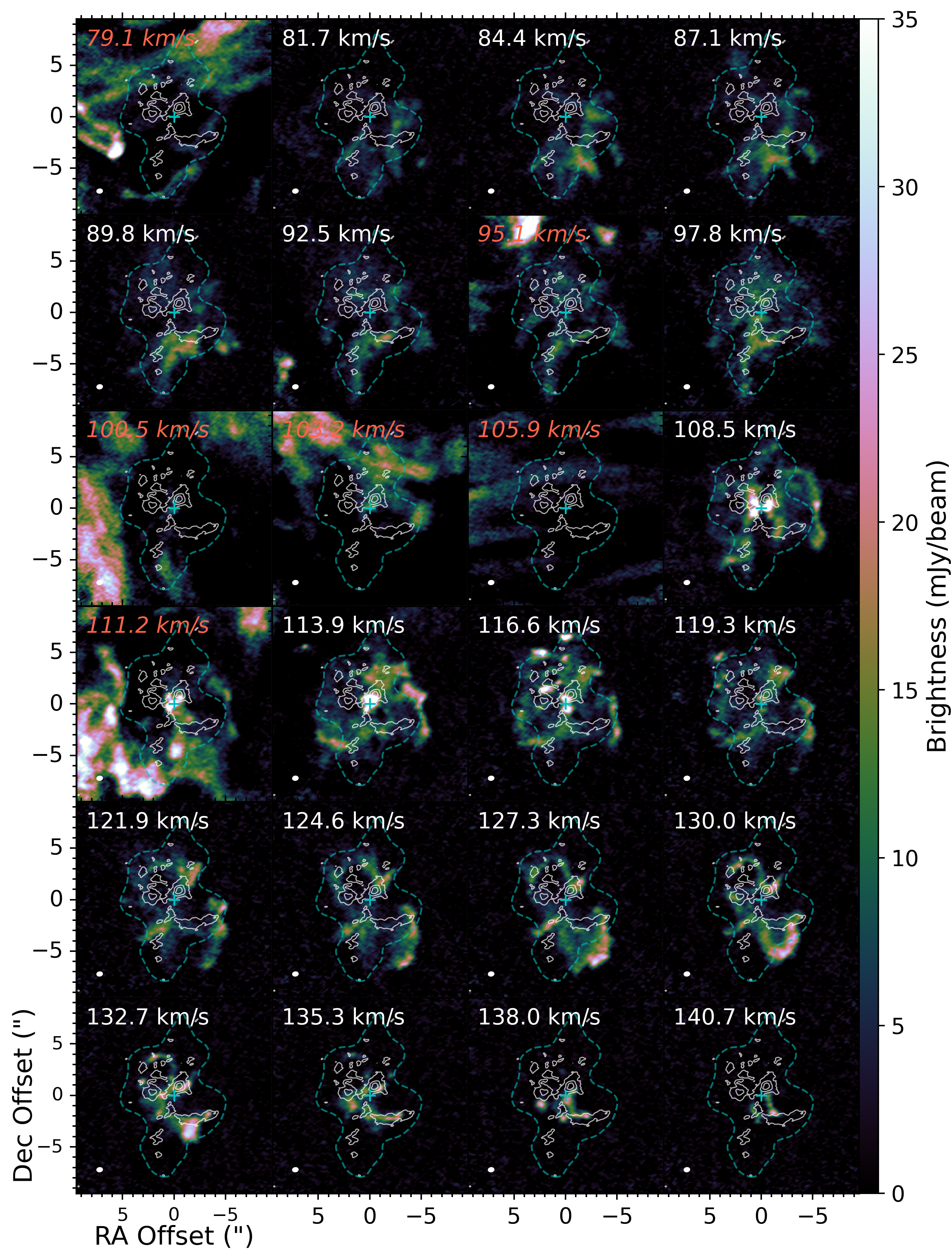}
    \caption{$^{12}$CO $J=2-1$ channel maps. The systemic velocity is $v_{sys}=109$\,\kms and the channel width is 1.34\,\kms. The white contours denote the 220\,GHz continuum emission at $5\sigma$, $10\sigma$, and $15\sigma$; the blue dashed contours represent $3\sigma$ for the smoothed extended continuum component; and the blue cross marks the stellar position. The ALMA beam is shown in the bottom left of each panel (in white). The velocities written in red italics indicate channel maps that are heavily affected by ISM contamination.}
     \label{fig:co_chmaps}
    \end{figure*}

    The individual channel maps for $^{12}$CO are shown in Fig.\ \ref{fig:co_chmaps}. We find that emission on the redshifted side of the line is characterised by two prominent bubble-like shapes extending in the NE and SW directions, as well as the bar structure discussed in Sect.\ \ref{sect:co}. In contrast, the blueshifted emission is more irregular, showing both smooth components as well as southward filamentary structures that do not converge to the stellar position at the highest relative velocities. Near the stellar velocity, the compact CO component appears, and we see bright emission from three confined clumps located at a radial distance of 1\arcsec\/ (5800\,au) from the stellar position.

\end{appendix}

\end{document}